\documentclass{aa}
\usepackage{graphicx}
\usepackage[]{natbib}
\bibpunct{(}{)}{;}{a}{}{,} 
\begin{document}
   \title{Diffraction-limited 800nm imaging with the\\
   2.56m Nordic Optical Telescope}

   \titlerunning{Diffraction-limited imaging with the NOT}

   \author{J.E. Baldwin \inst{1} \and R.N. Tubbs \inst{2} \and
           G.C. Cox \inst{3} \and C.D. Mackay \inst{2} \and
           R.W. Wilson \inst{4} \and M.I. Andersen \inst{5}
          }

   \offprints{R.N. Tubbs}

   \institute{Cavendish Astrophysics Group, Cavendish Laboratory,
              Madingley Road, Cambridge CB3 0HE, UK.\\
              email: jeb@mrao.cam.ac.uk
         \and
              Institute of Astronomy,
              Madingley Road, Cambridge CB3 0HA, UK.\\
              email: (RNT) rnt20@ast.cam.ac.uk (CDM) cdm@ast.cam.ac.uk
	 \and
              Nordic Optical Telescope,
              Apartado 474,
              E-38700 Santa Cruz de La Palma,
              Canarias, Spain\\
              email: cox@not.iac.es
         \and
	      Department of Physics,
	      University of Durham,
	      South Road,
	      Durham DH1 3LE, UK. \\
              email: r.w.wilson@durham.ac.uk
         \and
              Division of Astronomy,
              University of Oulu,
              P.O.BOX 3000
              FIN-90014 OULUN YLIOPISTO, Finland\\
              email: manderse@sun3.oulu.fi
             }

   \date{Received December 7, 2000 / Accepted January 23, 2000}

   \abstract{A quantitative assessment is presented of diffraction-limited stellar 
images with Strehl ratios of 0.25-0.30 obtained by selection of 
short-exposure CCD images of stars brighter than +6m at 810nm with the
Nordic Optical Telescope.
   }

   \maketitle

%
\section{Introduction}
The selection of the few sharpest images from a large dataset of short
exposures provides one method of obtaining high resolution images
through atmospheric seeing at ground-based telescopes. The selected
exposures can then be processed using one of the conventional speckle
techniques such as shift-and-add. Exposure selection has been
applied in many studies of the solar surface and for planetary imaging
but has not been extensively tested as a general tool for astronomical
imaging at optical wavelengths. Trials by \citet{dantowitz98} and
\citet{dantowitz00} using a
video camera on the 60-inch Mt.Wilson telescope have shown
its promise for reaching close to the diffraction limit. In view of
the technical difficulties of adaptive
optics at wavelengths shorter than 1 $\mu$m, it seems important to assess 
this alternative technique quantitatively.
In this paper we present such observations, showing the power of
the method and argue that new developments in low-noise, fast-readout
CCD's make it attractive for achieving diffraction-limited imaging in the
visible waveband for faint astronomical targets using apertures of about
2.5m.

The basis for the method is that the atmosphere behaves as a time-varying
random phase screen, with a power spectrum of irregularities characterised
by spatial and temporal scales \(r_{0}\) and \(t_{0}\), whose rms 
variations are larger than those of a well-adjusted and figured primary
mirror. Occasionally the combined phase
variations across the telescope aperture, \(\delta\), due to the
atmosphere and mirror, will be small
($<\sim1$radian). The corresponding image of a star will have a core which
is diffraction-limited with a Strehl ratio of \(\exp{(-\delta^{2})}\)
and angular resolution $1.22\lambda/D$ determined by the aperture diameter $D$.
\citet{fried78}
calculated the probability, \(P\), of ``lucky exposures'' having phase
variations less than 1 radian across an aperture diameter $D$ for
seeing defined by \(r_{0}\):
              \[P\simeq5.6\exp{\left(-0.1557\left(D/r_{0}\right)^{2}\right)}\]
This implies, for instance, that for an aperture
\(D=7r_{0}\), one
exposure in 350 would have a Strehl ratio greater than 0.37. For
\(D=10r_{0}\), the frequency of such good exposures falls to only one in
\(10^{6}\).
This
suggests that values of $D$ chosen as $7r_{0}$ may offer the best
compromise
between high angular resolution and frequency of occurrence.

The technique evidently requires a site with good seeing and a telescope
for which it is known that the errors in the mirror figure are small
compared with the atmospheric fluctuations on all relevant scales. The primary 
mirror should ideally match the maximum useful
aperture for this technique. If very good seeing is taken to be 0.5
arcsec, and a useful aperture as \(7r_{0}\), then D should be 1.4m at 500nm and
2.5m at 800nm. The Nordic Optical Telescope (NOT) matches these  criteria 
very closely. 

\section{Observations and Data Reduction}
Observations were made at the Cassegrain focus of the NOT on the nights of
2000 May 12 and 13. The camera was one developed for the JOSE programme of
seeing evaluation at the William Herschel Telescope \citep{stjacques97}. It comprised a
512x512 front illuminated frame-transfer CCD with 15 $\mu$m pixels run
by an AstroCam 4100 controller. The controller
allows windowing of the area of readout and variable pixel readout rates
up to 5.5MHz. The f/11 beam at the focus was converted to f/30 using a 
single achromat to give an image scale of 41 milliarcsec/pixel (25 pixels/
arcsec). This gives a good match to the full width to half maximum (FWHM) 
of the diffraction-limited image of 66 milliarcsec for a 2.56m telescope 
with 0.50m secondary obstruction at 800nm; good signal-to-noise sensitivity
per pixel is retained whilst the resolution is degraded only slightly
to 77 milliarcsec FWHM by the finite pixel size.

A filter centred at 810nm with a bandwidth of 120nm was used to define the
band. All of the exposures of stars were taken at frame rates higher than 
150Hz and without autoguiding to ensure that the temporal
behaviour of the periods of good seeing was adequately characterised. 

Observations of the stars listed in Table~\ref{target_list} were made with a variety of
frame formats and frame rates. Each run typically comprised between 5000 and
24000 frames over a period of 30-160s. Target stars, both single stars and
binaries, were chosen principally lying in the declination range $10\degr$-$20\degr$ and close
to the meridian, so that most of the data was taken at zenith angles
$<20\degr$. Zenith angles up to $50\degr$ were explored in later observations.
The effects of atmospheric dispersion became significant, since no corrective optics were employed.
   \begin{table}
      \caption[]{Observations}
         \label{target_list}
      \[
         \begin{array}{p{0.25\linewidth}p{0.3\linewidth}l}
            \hline
            \noalign{\smallskip}
            Target & Frame rate / Hz & FWHM / arcseconds \\
            \noalign{\smallskip}
            \hline
            \noalign{\smallskip}
            v656 Herculis & 185 & 0.49 \\
            $\epsilon$ Aquilae & 185 & 0.38 \\
            $\gamma$ Aquilae & 185 & 0.46 \\
            $\gamma$ Leonis & 159 & 0.57 \\
            $\gamma$ Leonis & 182 & 0.46 \\
            CN Bo\"otis & 152 & 0.62 \\
            $\zeta$ Bo\"otis & 152 & 0.74^{\mathrm{a}} \\
            $\zeta$ Bo\"otis & 152 & 0.75^{\mathrm{a}} \\
            $\alpha$ Herculis & 191 & 0.38 \\
            $\alpha$ Aquilae & 206 & 0.50 \\
            $\beta$ Delphini & 373 & 0.42 \\
            $\beta$ Delphini & 257 & 0.52 \\
            $\beta$ Delphini & 190 & 0.64 \\
            $\alpha$ Delphini & 180 & 0.41 \\
            $\alpha$ Delphini & 180 & 0.49 \\
            \noalign{\smallskip}
            \hline
         \end{array}
      \]
\begin{list}{}{}
\item[$^{\mathrm{a}}$] Includes some tracking error
\end{list}
\end{table}
The seeing was good, typically 0.5 arcsec (see
Table~\ref{target_list}), and the short exposure
images seen in real time clearly showed a single bright speckle at some 
instants. The full aperture of the NOT was therefore used for all the 
observations on both nights.

\begin{figure}
\resizebox{\hsize}{!}{\includegraphics{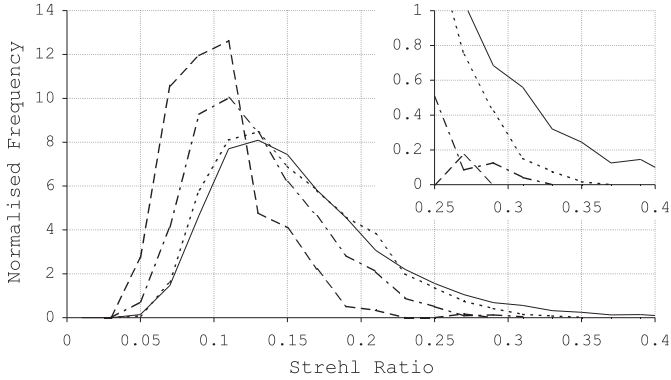}}
\caption{
Histogram of the Strehl ratios calculated for 5950 exposures of \object{$\epsilon$
Aquilae} taken over a period of $\sim$30 seconds. The dotted curve shows 
the measured distribution of Strehl ratios for the individual exposures. 
The solid line is for a
computer simulation of Kolmogorov turbulence with an aperture diameter of
$6.5r_{0}$. The dot-dashed curve shows the Strehl ratios which result after
averaging together groups of five consecutive exposures of
\object{$\epsilon$ Aquilae} \emph{without}
correction for image motion. If groups of 20 consecutive
exposures are averaged together the resulting images typically have
substantially lower Strehl ratios (dashed line). For this case only
one of the averaged images gives a Strehl greater than 0.25.
}
\label{strehl_histogram}
\end{figure}

The analysis of the data was carried out in stages:
\begin{enumerate}
\item Some runs in which the detector was saturated during the periods of
best seeing and a small number of
misrecorded frames in runs taken at the fastest frame rates were excluded 
from further analysis.
\item For each run, an averaged image was obtained by summing all the
remaining frames. The mean sky level was measured from a suitable area of the image
and subtracted before determining the FWHM of the
seeing disk (Table~\ref{target_list}) and the total stellar flux. 
\item Each frame was interpolated to give a factor of 4 times as many
pixels in each coordinate using sinc interpolation to overcome
problems of sampling in the images.  
\item The peak brightness and position of the brightest pixel in each
frame was measured and stored.
\item The Strehl ratio for each frame was derived using the peak
brightness above the sky level measured for that frame,
scaled by the total flux from the star and a geometrical factor which
relates the pixel scale to the theoretical diffraction response of the NOT.
\item A ``good'' image was then obtained by taking only those frames with a Strehl
ratio greater than some chosen value, shifting the images so that the peak
pixels were superposed and adding the frames.
\end{enumerate}
Fig.~\ref{strehl_histogram} shows a histogram of the Strehl ratios
derived following 1-5 above for individual 5.4 ms exposures in a 32
second run on the star \object{$\epsilon$ Aquilae}. The substantial
improvement in the Strehl of the shift-and-add image and in limiting sensitivity provided by image
selection is clear from the spread of the Strehl values.
For comparison, a histogram is also plotted of the Strehl ratios obtained 
for images from simulations of 10,000 realisations of atmospheric 
irregularities with a Kolmogorov turbulent spectrum over a circular 
aperture \(6.5r_{0}\) in diameter. The close agreement between theory and 
observation suggests that the conditions corresponded to \(6.5r_{0}\) =
2.56m at 810nm for this run. The FWHM of the averaged image from this run 
was 0.4 arcsec, corresponding to the telescope aperture being only \(5.8r_{0}\)
for a Kolmogorov spectrum of turbulence. The discrepancy is partly
explained by the actual image motion being only 60\% of that for a Kolmogorov
spectrum, implying an outer scale of turbulence of about 30m.

The timescale on which changes take place in the best images is an
important question, setting a limit to the maximum frame exposure time which
can be used in this technique, and hence to the faintest limiting
magnitude achievable with a given camera. Histograms of the Strehl ratios
obtained after averaging together groups of 5 successive frames without 
correction for image motion are included in Fig.~\ref{strehl_histogram}. This shows that 
exposures as long as 30ms could be used with only a small reduction in 
the observed Strehl values. The Strehls are reduced 
substantially if 20 exposures are averaged together. The choice of best 
exposure time depends on the intended application.
%
\section{Results}
\begin{figure}
\resizebox{\hsize}{!}{\includegraphics{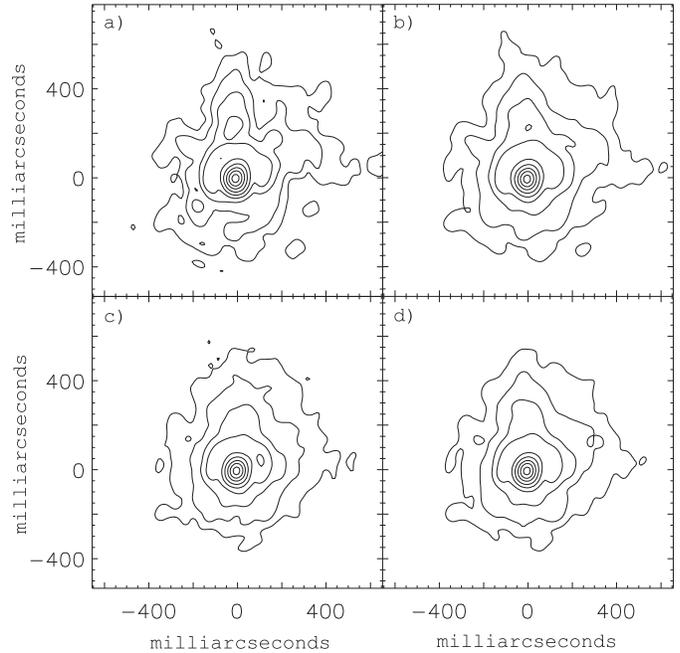}}
\caption{
\textbf{a-d.} Image quality of \object{$\epsilon$ Aquilae} using differing criteria for exposure 
selection from a 32s run. Contour levels are at 1, 2, 4, 8, 16, 30, 50, 
70,
90\% peak intensity.
}
\parbox{\linewidth}
{
\newcounter{alpha_count}
\begin{list}{\textbf{\alph{alpha_count})}}{\usecounter{alpha_count}
	\setlength{\rightmargin}{\leftmargin}}
\item The single best 108ms exposure. Strehl~=~0.29.
\item 12 exposures of 27ms duration combined using shift-and-add.
Strehl~=~0.26.
\item 20 individual 5.4ms exposures from \object{$\epsilon$ Aquilae}
taken from widely separated time periods, shifted and added
together. Strehl~=~0.30.
\item The 60 individual 5.4ms exposures of \object{$\epsilon$ Aquilae} with the highest Strehl
ratio, shifted and added together. Strehl~=~0.30.
\end{list}
}
\label{eaql_contour}
\end{figure}

Data from the run on \object{$\epsilon$ Aquilae} analysed in several
different ways illustrate the
possible strategies in practice. Fig.~\ref{eaql_contour}a shows a 108ms exposure, 
formed by the summation \emph{without} shifting of 20 consecutive 5.4 ms 
exposures, during a single period of good atmospheric conditions. The FWHM 
of the central core in the image is 79$\times$95 milliarcseconds, with a 
Strehl ratio of 0.29. Fig.~\ref{eaql_contour}b was constructed from 12
exposures of 27ms duration, each made by 
summing \emph{without} shifting five 5.4ms exposures. These 12 exposures
were shifted and added to give a final FWHM of 81$\times$96
milliarcseconds and a Strehl ratio of 0.26. Single 5.4ms exposures from 
20 widely separated time periods were selected for Fig.~\ref{eaql_contour}c. Each of the
constituent exposures had a Strehl ratio greater than 0.27, resulting
in an image core with 80$\times$93 milliarcseconds FWHM and a Strehl of 0.30
after shifting and adding. For Fig.~\ref{eaql_contour}d the
exposures with the highest 1\% of Strehl ratios are selected, giving a
final image with FWHM of 79$\times$94 milliarcseconds and Strehl
ratio of 0.30. All of these procedures give images with
high Strehl ratios. There is, however, a trade-off between the higher
sensitivity for faint reference stars achieved by the long sequence in
Fig.~\ref{eaql_contour}a and the smoothness of the halo due to
averaging many atmospheric configurations as in Fig. 2. b-d.
For the remainder of this letter we have used the best 1\%
of exposures in any observation to produce the final image.

\begin{figure}
\resizebox{\hsize}{!}{\includegraphics{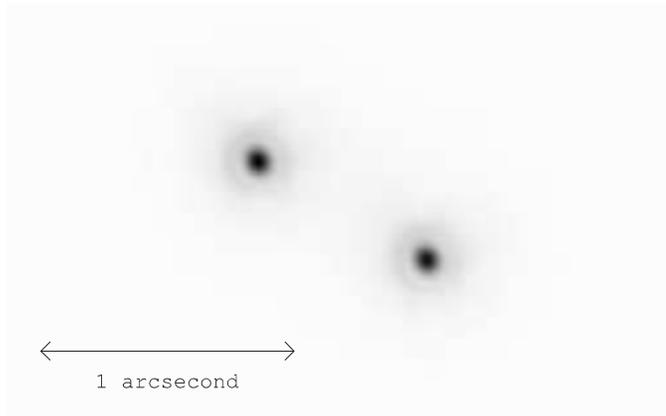}}
\caption{
The best 1\% of exposures of \object{$\zeta$ Bo\"otis}, shifted and added.
}
\label{zboo_bitmap}
\end{figure}

\begin{figure}
\resizebox{\hsize}{!}{\includegraphics{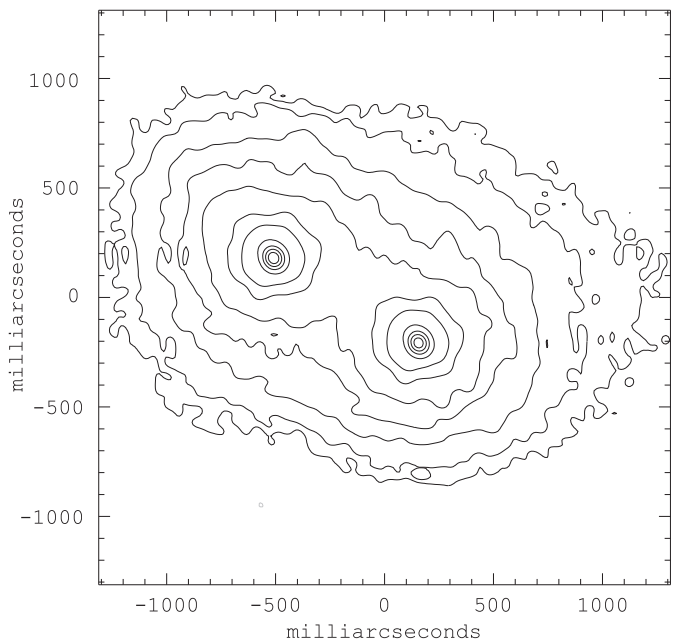}}
\caption{
Contour plot of the image of \object{$\zeta$ Bo\"otis} in
Fig.~\ref{zboo_bitmap}. Contour levels
at 0.1, 0.2, 0.5, 1, 2, 5, 10, 20, 40, 60, 80\% peak intensity.
}
\label{zboo_contour}
\end{figure}

Fig.~\ref{zboo_bitmap} shows an example image of \object{$\zeta$
Bo\"otis} generated by selecting exposures from a
dataset of 23200 frames. The image shows a diffraction-limited
central peak (Strehl = 0.26, FWHM = 83$\times$94 milliarcseconds) and first Airy
ring superposed on a faint halo for each star. 300
milliarcseconds from the component stars the
surface brightness of the halo reaches only 2\% of that in the stellar
disks. The 232 frames selected came from a range of
different epochs, helping to reduce the level of fluctuation around
the stars. The high dynamic range of the technique is evident from the
contour plot of the same image (Fig.~\ref{zboo_contour}). The fluctuations in
the halo reach only 0.1\% of the peak brightness 700 milliarcseconds
from the stars. The magnitude difference between the two components in
this image is 0.048 $\pm$ 0.005, in good agreement with \citet{hipparcos97}.

\begin{figure} \resizebox{\hsize}{!}{\includegraphics{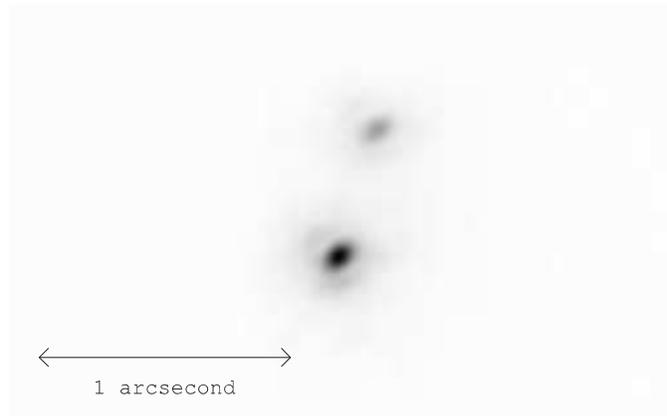}}
\caption{
The best 1\% of exposures of \object{$\beta$ Delphini}, shifted and added. 
}
\label{bdel_bitmap}
\end{figure}

Fig.~\ref{bdel_bitmap} shows the result of a similar selection of the 1\% of 
images with the best 
Strehl ratios from a dataset of 7000 short-exposure CCD images of 
\object{$\beta$ Delphini}. In this case 
the zenith angle of the observation was 50 degrees and the images are 
blurred by 100milliarcsec due to atmospheric dispersion over the 120nm
bandpass of the filter, reducing the Strehl ratio of the final image to 
0.25. The magnitude difference between the components is
$\Delta$$M=1.070\pm0.005$. This value is in good agreement with those of 
\citet{barnaby00} of $1.071\pm0.004$ at 798nm and $1.052\pm0.010$ at 
884nm made using a 1.5m telescope. The images shown of
\object{$\epsilon$ Aquilae}, \object{$\zeta$
Bo\"otis} and \object{$\beta$ Delphini} are representative of those for all
the stars listed in Table~\ref{target_list} with regards to Strehl
ratio and core FWHM.
%
\section{Conclusions and Future Prospects}
The observations described here show that selection of short exposure images can
reliably provide diffraction-limited images with Strehl ratios of 0.25-0.30 at 
wavelengths as short as 0.8 microns with 2.5m telescopes. The
images are similar in core angular resolution and Strehl ratio to the
highest resolution images from
adaptive optics at wavelengths shorter than 1 $\mu$m \citep{graves98}.
The faintest stars used in these trial observations were of magnitude
 +6. The readout noise of
the present camera ($\sim100e^{-}$) would have set a limiting magnitude of 
+11.5 if 30ms exposures had been used. Current development of CCD 
detectors with effectively zero readout noise and higher quantum
efficiency \citep{mackay01} will provide
important advantages in the use of the exposure selection method in the near future. The 
limiting magnitude for reference stars is expected to be fainter than +15.5, and
fainter than +23 for unresolved objects in the same isoplanatic patch. In cases 
where the seeing is dominated primarily by one turbulent atmospheric 
layer at a height $H$, the diameter of the isoplanatic patch at the times 
of the good selected exposures will be $7r_{0}/H$ or say 2.5m/5km = 1.7 
arcmin. There would then be 
$>80\%$ sky coverage by usable reference stars under 0.5 
arcsec seeing conditions for 2.5m telescopes.
%
\begin{acknowledgements}
The Nordic Optical Telescope is operated on the island of La Palma
jointly by Denmark, Finland, Iceland, Norway and Sweden, in the
Spanish Observatorio del Roque de los Muchachos of the Instituto de
Astrofisica de Canarias.
\end{acknowledgements}
%

\bibliographystyle{apj} 
\bibliography{bibtex/je_baldwin}
\end{document}